\documentclass[prb,twocolumn,letterpaper,superscriptaddress,footinbib]{revtex4}
\usepackage[utf8x]{inputenc}
\usepackage{graphicx}
\usepackage{dcolumn}
\usepackage{bm}
\usepackage{subfigure}
\usepackage{mathrsfs}
\usepackage{amsmath,amsfonts,amssymb,latexsym}
\usepackage[usenames]{color}
\usepackage{epstopdf}

\newcommand{\cE}{\mathcal E}

\newcommand{\tr}{\text{Tr}}

\definecolor{BrickRed}{cmyk}{0,0.89,0.94,0.28}
\definecolor{MidnightBlue}{cmyk}{0.98,0.13,0,0.43}
\definecolor{DarkGreen}{rgb}{0,0.7,0.1}

\definecolor{RedViolet}{cmyk}{0.07,0.90,0,0.34}
\definecolor{SeaGreen}{cmyk}{0.69,0,0.50,0}
\definecolor{FireOrange}{rgb}{1.,.294,.247}

\begin{document}
\title{Spin texture on the Fermi surface of tensile strained HgTe}
\author{Saad Zaheer}
\affiliation{Department of Physics and Astronomy, University of Pennsylvania, Philadelphia, Pennsylvania 19104-6396, USA}
\author{S. M. Young}
\affiliation{The Makineni Theoretical Laboratories, Department of Chemistry, University of Pennsylvania, Philadelphia, Pennsylvania 19104-6323, USA}
\author{D. Cellucci}
\affiliation{Department of Physics and Astronomy, University of Georgia, Athens, Georgia 30602, USA}
\author{J. C. Y. Teo\footnote{Present address: Department of Physics, University of Illinois at Urbana-Champaign, Urbana, Illinois 61801-3080, USA}}
\affiliation{Department of Physics and Astronomy, University of Pennsylvania, Philadelphia, Pennsylvania 19104-6396, USA}
\author{C. L. Kane}
\affiliation{Department of Physics and Astronomy, University of Pennsylvania, Philadelphia, Pennsylvania 19104-6396, USA}
\author{E. J. Mele}
\affiliation{Department of Physics and Astronomy, University of Pennsylvania, Philadelphia, Pennsylvania 19104-6396, USA}
\author{Andrew M. Rappe}
\affiliation{The Makineni Theoretical Laboratories, Department of Chemistry, University of Pennsylvania, Philadelphia, Pennsylvania 19104-6323, USA}
\date{\today}

\begin{abstract}
We present \emph{ab initio} and ${\bf k\cdot p}$ calculations of the spin texture on the Fermi surface of tensile strained HgTe, which is obtained by stretching the zincblende lattice along the (111) axis. Tensile strained HgTe is a semimetal with pointlike accidental degeneracies between a mirror symmetry protected twofold degenerate band and two nondegenerate bands near the Fermi level. The Fermi surface consists of two ellipsoids which contact at the point where the Fermi level crosses the twofold degenerate band along the (111) axis. However, the spin texture of occupied states indicates that neither ellipsoid carries a compensating Chern number. Consequently, the spin texture is locked in the plane perpendicular to the (111) axis, exhibits a nonzero winding number in that plane, and changes winding number from one end of the Fermi ellipsoids to the other. The change in the winding of the spin texture suggests the existence of singular points. An ordered alloy of HgTe with ZnTe has the same effect as stretching the zincblende lattice in the (111) direction. We present \emph{ab initio} calculations of ordered $\rm Hg_{x}Zn_{1-x}Te$ that confirm the existence of a spin texture locked in a 2D plane on the Fermi surface with different winding numbers on either end. 
\end{abstract}
\maketitle

\begin{figure}[t]
{
     \includegraphics [width=3in]{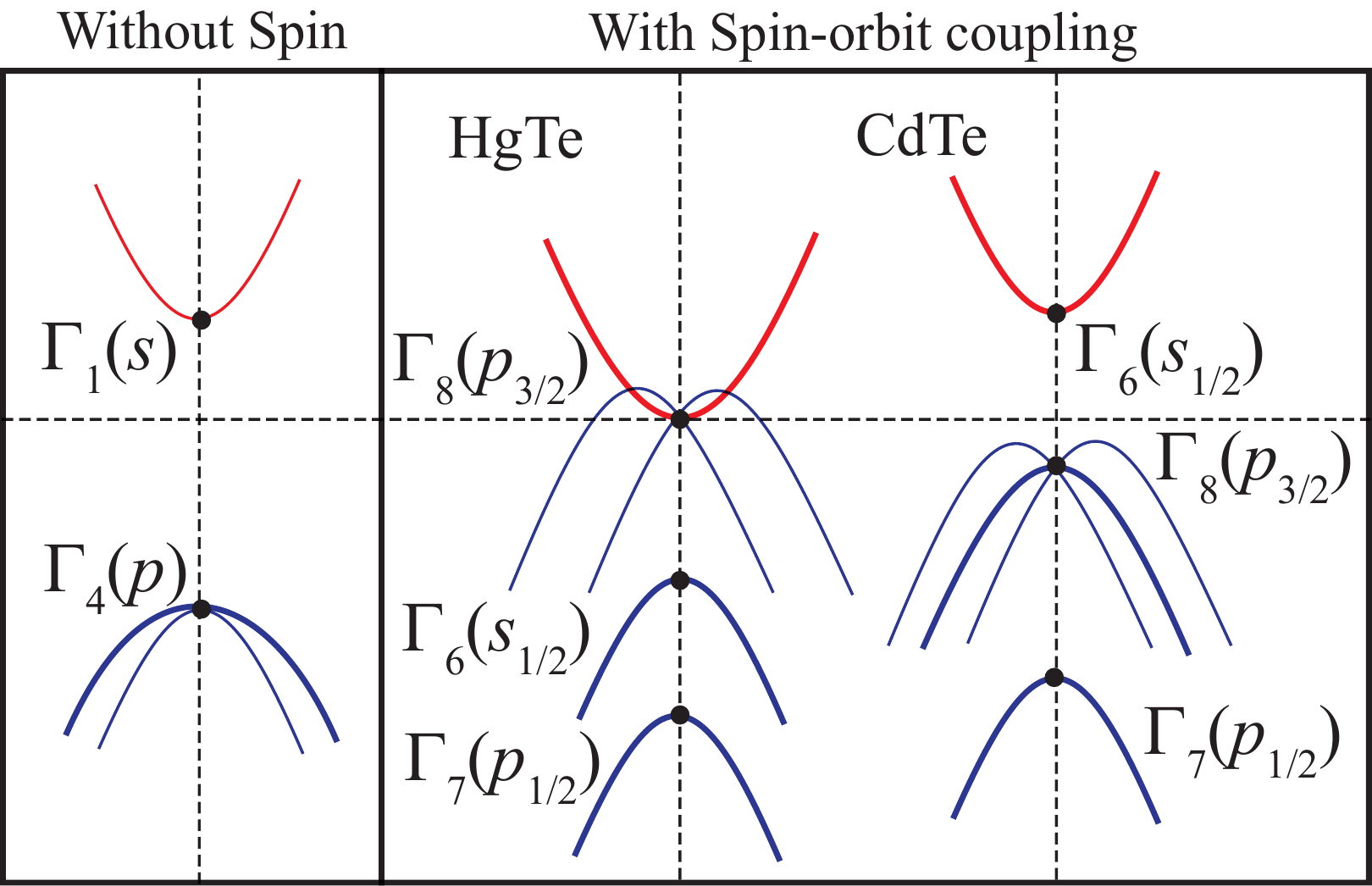}
}

\caption{{Schematic illustration of band inversion in HgTe due to strong spin-orbit coupling. Twofold degenerate bands are drawn as thick lines, whereas nondegenerate bands are drawn as thin lines. Left: Without spin, HgTe/CdTe have an $s$-type conduction band and a $p$-type valence band. Middle: The spin-orbit coupling in HgTe is strong enough to push the $s_{1/2}$-type representation $\Gamma_6$ below the $p_{3/2}$-type representation $\Gamma_8$. Right: Normal ordering of states in CdTe, which has a weaker spin-orbit coupling as compared with HgTe. The $ s_{1/2}$-type representation $\Gamma_6$ appears above the $ p_{3/2}$-type representation $\Gamma_8$, as expected from the spinless case.}}\label{inversion}
\end{figure}

\begin{figure*}[t]
{
\subfigure[]{\includegraphics [width=3in]{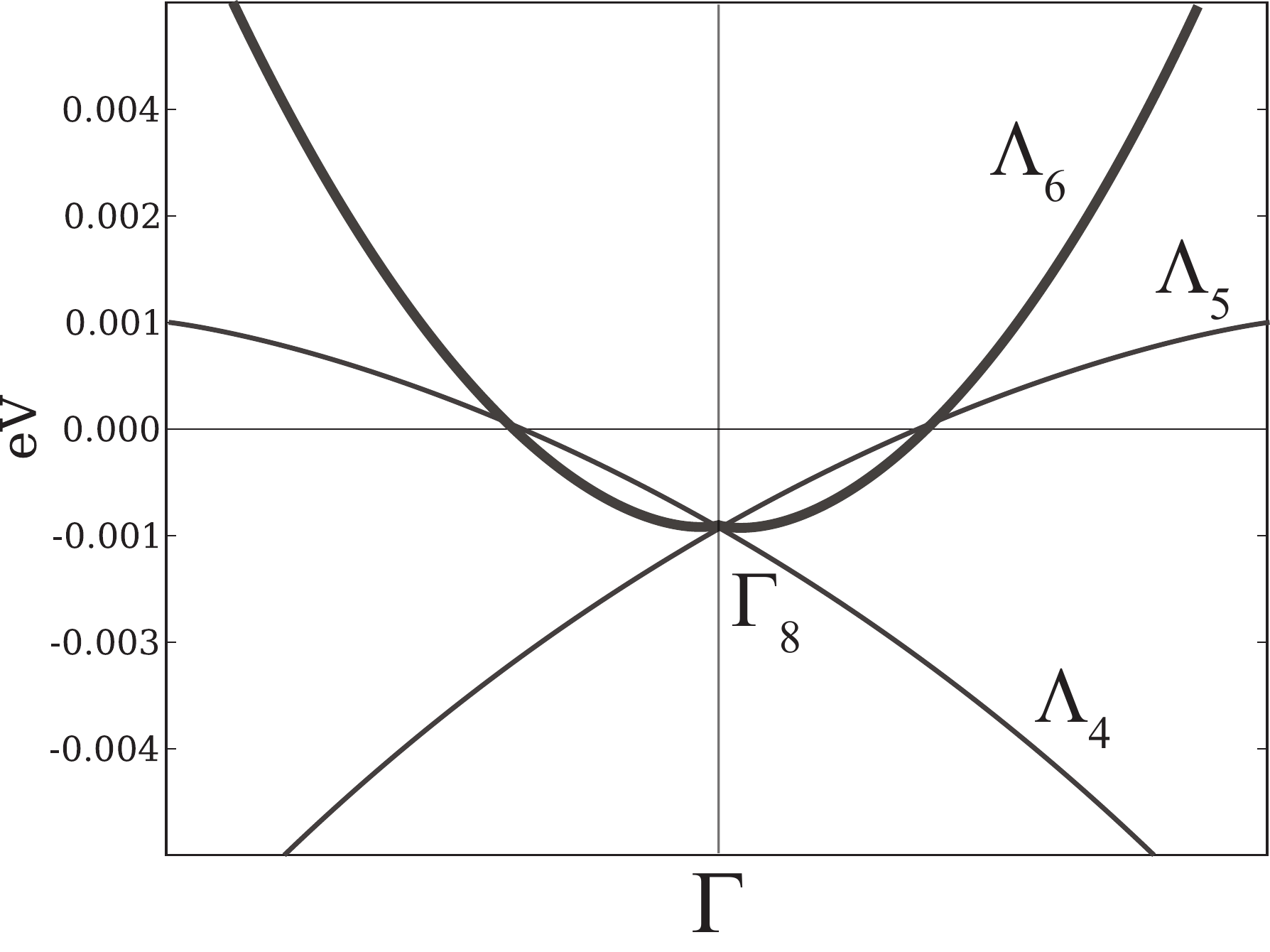}\label{fig:kpgamma}}
\subfigure[]{ \includegraphics [width=3.5in]{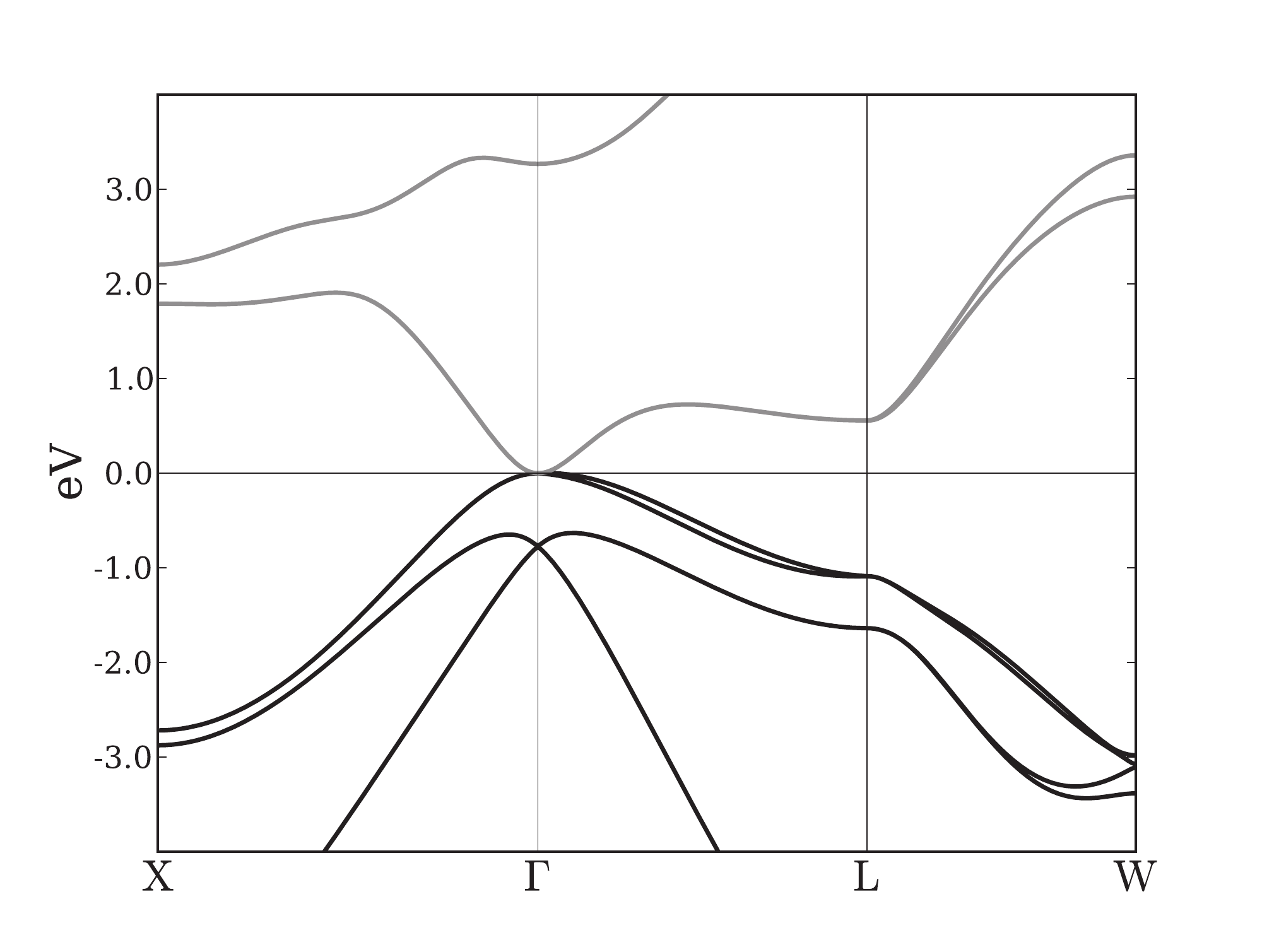}\label{fig:ref}}
}

\caption{{Band structure of HgTe. ~\subref{fig:kpgamma} Energy bands near the Fermi level as a function of momentum along the (111) axis. At $\Gamma$ the four degenerate states span the irreducible representation $\Gamma_8$ of the group $T_d$. $\Lambda_{4,5,6}$ are irreducible representations of the group $ C_{3v}$. The twofold degenerate (solid) band labeled $\Lambda_6$ disperses quadratically to lowest order, whereas the nondegenerate bands $\Lambda_4$ and $\Lambda_5$ disperse linearly. There are threefold accidental degeneracies at ${\bf k} = \pm 0.004(1,1,1)$ because the bands $\Lambda_4$ and $\Lambda_5$ have negative mass, whereas $\Lambda_6$ has positive mass. ~\subref{fig:ref} Band structure of HgTe as a function of crystal momentum across the entire Brillouin zone. The valence and conduction bands are gapped everywhere except near $\Gamma$.}}
\end{figure*}

Mercury telluride (HgTe) is a zero band-gap zincblende semiconductor with an inverted ordering of states at $\Gamma$~\cite{springer}, where the band gap and inversion can be controlled by alloying with CdTe. HgTe and CdTe are both zincblende materials, with the crucial difference that the stronger spin-orbit coupling of Hg causes the states at $\Gamma$ to invert relative to each other [Fig.~\ref{inversion}]. Therefore HgTe/CdTe quantum wells exhibit the quantum spin Hall effect~\cite{Bernevig06p1757}. The electronic properties of alloys of HgTe have been studied extensively for many decades. Recently there has been a surge of interest in so-called topological (Weyl) semimetals: three dimensional materials which exhibit pointlike degeneracies (Weyl points~\footnote{The Weyl Hamiltonian describes two linearly dispersing bands around a degenerate point. In simplest terms, it can be written as $\hat{H}({\bf k}) = k_x\sigma_x+ k_y\sigma_y+k_z\sigma_z$.}) between bulk conduction and valence bands and exhibit topologically protected surface modes~\cite{Burkov11p127205,Vishwanath11p205101,Yang11p075129}. These topological Weyl points are low symmetry descendants of a so-called Dirac point, which is described by a massless four-band Dirac-like Hamiltonian~\cite{Young12p140405}. The unique feature of a Dirac semimetal is a Fermi surface that surrounds discrete Dirac points, around which all bands disperse linearly in all directions. It turns out that HgTe is \emph{almost} a Dirac semimetal: at $\Gamma$ two conduction states are degenerate with two valence states; all four bands disperse linearly along the (110) and (100) axes; however, mirror and time reversal symmetry guarantee that along the (111) axis, two of the four bands must disperse \emph{quadratically}~[Refs.~\onlinecite{Young12p140405, Dresselhaus55p580} and Fig.~\ref{fig:kpgamma}]. 

Figure~\ref{fig:kpgamma} shows another interesting feature of the bands at $\Gamma$. In addition to the symmetry guaranteed fourfold degeneracy, there are \emph{additional} pointlike accidental degeneracies along the (111) axis between two conduction states and one valence state. The rest of the band structure is completely gapped, so the Fermi surface is confined to a small region around $\Gamma$~[Fig.~\ref{fig:ref}]. It is natural to ask what perturbations could turn HgTe into a Dirac semimetal. Since the quadratic dispersion of bands near $\Gamma$ is enforced by mirror and time reversal symmetry, any perturbation must break either one of these symmetries. A time reversal symmetry breaking Zeeman field breaks at least one rotation symmetry of the tetrahedral group splitting the fourfold degeneracy at $\Gamma$ so the system either gaps or develops additional Fermi surface. Breaking mirror symmetry maintains the fourfold degeneracy at $\Gamma$, but the Dirac point develops a nonzero Chern number due to the threefold rotation symmetry of zincblende. This inevitably leads to additional Fermi surface as well. Hence, zincblende materials cannot be Dirac semimetals~\cite{Young12p140405}. A third option is to break, for instance, a fourfold rotation symmetry by straining the lattice. Such a procedure will split the fourfold degeneracy at $\Gamma$ and will either gap the band structure completely or else shift the accidental degeneracy away from $\Gamma$ along the (111) axis.

\begin{figure*}[t]
{
\subfigure[]{     \includegraphics [width=3in]{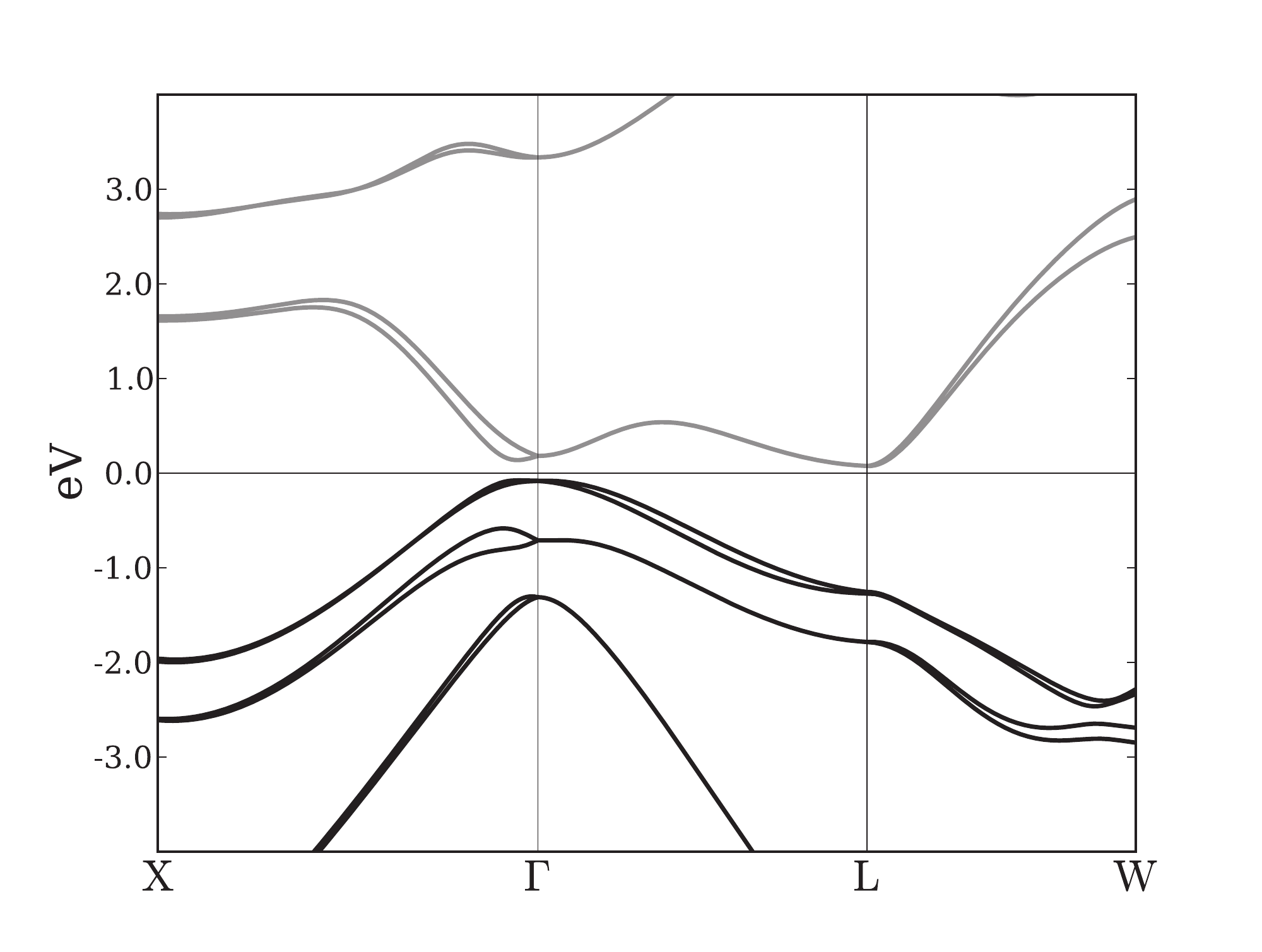}\label{fig:neg}}
\subfigure[]{     \includegraphics [width=3in]{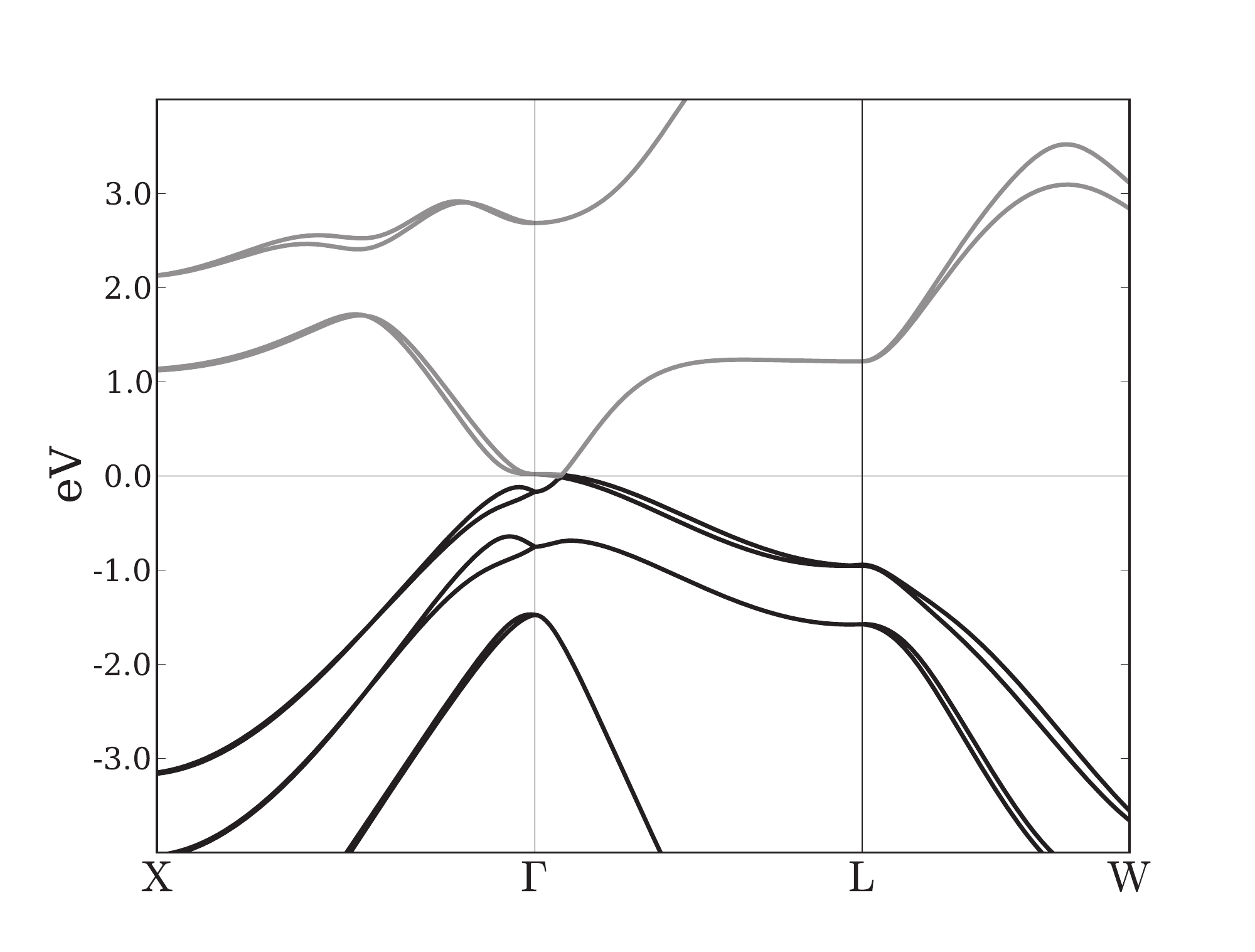}\label{fig:pos}}
}
\caption{{~\subref{fig:neg} Band structure of HgTe under compressive strain along the (111) axis. The fourfold degenerate representation $\Gamma_8$ is split, and the material becomes a topological insulator~\cite{Kane07p045302}.~\subref{fig:pos} Band structure of HgTe under tensile strain along the (111) axis. Although $\Gamma_8$ splits under tensile strain as well, the accidental degeneracy along the (111) axis shifts away from $\Gamma$ and the material remains semimetallic.}}\label{hgtebands}
\end{figure*}

Reference~\onlinecite{Kane07p045302} predicted that compressive strain along the (111) axis will gap HgTe into a topological insulator, since the states at $\Gamma$ are already inverted [Fig.~\ref{fig:neg}]. The inverted ordering of bands at $\Gamma$ is a direct consequence of spin-orbit coupling. To understand how it comes about, we compare the ordering of states in HgTe with and without spin~[Fig.~\ref{inversion}]. At the Fermi level in spinless HgTe, the valence states span a $p$-type representation $\Gamma_4$ of space group 216. Slightly above $\Gamma_4$ in energy, there exists an $s$-type band which belongs to the symmetric representation $\Gamma_1$. When spin-orbit coupling is introduced, the representation $\Gamma_4$ splits into a $p_{3/2}$-type representation $\Gamma_8$ and a $p_{1/2}$-type representation $\Gamma_7$, whereas $\Gamma_1$ turns into an $s_{1/2}$-type representation $\Gamma_6$. Since Hg is a heavy atom, its contribution to the spin-orbit coupling is strong enough to push the $s_{1/2}$-type band $\Gamma_6$ below $\Gamma_8$, resulting in an inverted band structure. This is in contrast with a normal ordered zincblende semiconductor (e.g. CdTe), where $\Gamma_6$ appears above $\Gamma_8$ in energy. Reference~\onlinecite{Bernevig06p1757} demonstrated that if $\Gamma_8$ is split and the inversion of $\Gamma_6$ relative to the split band $\Gamma_8$ is controlled by a ``mass'' parameter, the insulating phases that ensue have a non-trivial $\mathbb{Z}_2$ index $+1$ in the inverted regime, versus a trivial index $0$ in the normal regime. The splitting of $\Gamma_8$ was achieved in a HgTe/CdTe quantum well structure, while the tuning of $\Gamma_6$ above or below the split representation $\Gamma_8$ was controlled by the well thickness, leading to the first experimental observation of the quantum spin Hall effect~\cite{Konig07p766}. 

In this paper, we show that whereas compressive strain along the (111) axis gaps the system entirely~[Fig.~\ref{hgtebands}\subref{fig:neg}], tensile strain along the same axis shifts the accidental degeneracies further along the $(111)$ axis after splitting $\Gamma_8$~[Fig.~\ref{hgtebands}\subref{fig:pos}]. Tensile strained HgTe remains a semimetal where the dispersion of bands around the accidental degeneracies is linear in all directions~[Fig.~\ref{fermisurf}]. But there is a caveat. Each accidental degeneracy involves \emph{three} states so it is neither a Dirac point nor a Weyl point. As discussed earlier, this is due to the presence of mirror symmetry about a plane that contains the (111) axis. If we could break the mirror symmetry, the twofold degenerate band in Fig.~\ref{fig:kpgamma} would split and the system will develop two Weyl points [Fig.~\ref{kdotpbands}].

A Weyl point carries a nonzero Chern number (charge) $\pm 1$ and cannot be annihilated unless brought in contact with another Weyl point of opposite charge. In other words, the spin texture of valence states around a Weyl point has nonzero divergence; if interpreted as a magnetic field, the spin texture would correspond to a magnetic monopole with charge $\pm 1$. However mirror symmetry constrains all points on its mirror plane to have zero Chern number. Hence Weyl points cannot exist on mirror planes, and so the integral of the normal component of spins on a closed surface that encloses points on a mirror plane must be zero. 

This has important implications for the Fermi surface of tensile strained HgTe~[Fig.~\ref{fermisurf}]. The two Fermi ellipsoids contact at a point due to mirror symmetry. They can be separated only if mirror symmetry is broken. Consequently their corresponding spin textures will develop a nonzero divergence. We ask what is the nature of the spin texture on the ellipsoidal Fermi surface \emph{before} mirror symmetry is broken. It turns out that spin texture on both ellipsoids is locked in a plane perpendicular to the (111) axis and exhibits a nonzero winding number about that axis. Furthermore the winding number changes as one sweeps across the Fermi ellipsoids from one end to the other and inevitably exhibits spin singularities on the Fermi surface where the winding number must change~[Fig.~\ref{spintexture}]. 

We alloy HgTe with ZnTe by replacing half the Hg atoms with Zn atoms. The size mismatch between Hg and Zn causes the lattice to relax so that the Zn and Hg atoms move off center towards the Te atoms. This has the same effect as tensile strain along the (111) axis in zincblende~[Fig.~\ref{hgzntebands}]. Indeed, \emph{ab initio} calculations indicate that the spin texture on the Fermi surface of $\rm Hg_{x}Zn_{1-x}Te$ ($x=0.5$) is locked in a two dimensional plane and exhibits spin singular points identical to tensile strained HgTe~[Fig.~\ref{spinsurf}]. However it is likely that Density Functional Theory (DFT) overestimates~\cite{Svane11p205205} the amount of band inversion for $x=0.5$ given that $\rm Hg_{x}Cd_{1-x}Te$ is a band insulator for $x<0.8$~\cite{Scott69p4077}. Nonetheless, for small enough concentrations of Zn, this symmetry protected spin texture may indeed by realized physically. It is difficult to span the full range of intermediate compositions within DFT to make a prediction of the composition where the inversion occurs. 

In Section~\ref{secfermi}, we describe the Fermi surface of tensile strained HgTe. In Section~\ref{kdotptheory}, we derive a simple $\bf k \cdot p$ model to understand band dispersion in tensile strained HgTe. In Section~\ref{abinitio}, we derive the spin texture at the ${\bf k \cdot p}$ level and provide \emph{ab initio} calculations of $\rm Hg_{0.5}Zn_{0.5}Te$ in support of the predictions of the $\bf k \cdot p$ theory. Finally we conclude with some brief remarks. 

\section{Fermi Surface}\label{secfermi}
The band structure of HgTe has been of considerable interest in recent years because HgTe alloyed with CdTe exhibits a tunable direct band-gap from 0 to 1.5 eV at $\Gamma$. We focus our attention to the four dimensional irreducible representation $\Gamma_8$ at the Fermi level. The point group elements of the little group at $\Gamma$ belong to $T_d$ (the tetrahedral group with mirror planes). Since the zincblende lattice lacks inversion symmetry, the matrix elements $\langle \psi_i |\hat{{\bf p}}| \psi_j\rangle$, where $\{| \psi_i \rangle\}$ span $\Gamma_8$, are non-zero. Therefore, the four bands degenerate at $\Gamma$ can, in principle, disperse linearly in all directions around $\Gamma$~\cite{Young12p140405}. However, along the $(111)$ axis, $\Gamma_8$ splits into a two dimensional representation $\Lambda_6$ and two one dimensional representations $\Lambda_4$ and $\Lambda_5$ of the little group along that axis. Since $\Gamma$ is invariant under time reversal symmetry, $\Lambda_6$ is constrained to be flat, whereas $\Lambda_4$ and $\Lambda_5$ must disperse with non-zero slope. Along other directions such as (110) and (100), $\Gamma_8$ splits into either four singlets or two doublets, and none of the bands are constrained to be flat to linear order in ${\bf k}$~\cite{Young12p140405, Dresselhaus55p580}. This is why the four band model describing states $\{|\Gamma_8,3/2\rangle, |\Gamma_8,-3/2\rangle,|\Gamma_6,1/2\rangle,|\Gamma_6,-1/2\rangle \}$ in Ref.~\onlinecite{Bernevig06p1757} was written as a Dirac Hamiltonian in \emph{two} dimensions. 

It turns out that the bands belonging to $\Lambda_4$ and $\Lambda_5$ have negative mass [Fig.~\ref{fig:ref}]. Therefore, to $\mathcal{O}({\bf k}^2)$ and above, the bands labeled $\Lambda_4$ and $\Lambda_5$ must bend back below the Fermi level and cross $\Lambda_6$ somewhere along the $(111)$ axis. High resolution \emph{ab initio} calculations confirm that these degeneracies indeed occur very close to $\Gamma$ at $~\pm0.004(1,1,1)$ [Fig.~\ref{fig:kpgamma}]. Since this occurs so close to $\Gamma$, the Fermi surface of HgTe is only effectively pointlike.

Under tensile strain along the $(111)$ axis, the representation $\Gamma_8$ splits, because the tetrahedral group $ T_d$ reduces to the trigonal group $C_{3v}= \{1, x,x^2,y, xy, x^2y\}$. Here $x$ corresponds to threefold rotation symmetry about the $(111)$ axis while $\{y, xy, x^2y\}$ represent the three mirror planes that contain the (111) axis and are rotated by $2\pi/3$ with respect to each other. Since tensile strain preserves the little group along the line from $\Gamma$ to L, the representation $\Lambda_6$ does not split. Therefore, the accidental degeneracies between $\Lambda_4/ \Lambda_5$ and $\Lambda_6$ only shift further away from $\Gamma$, and HgTe remains a semimetal. 

\begin{figure}[t]
{

    \includegraphics [width=3in]{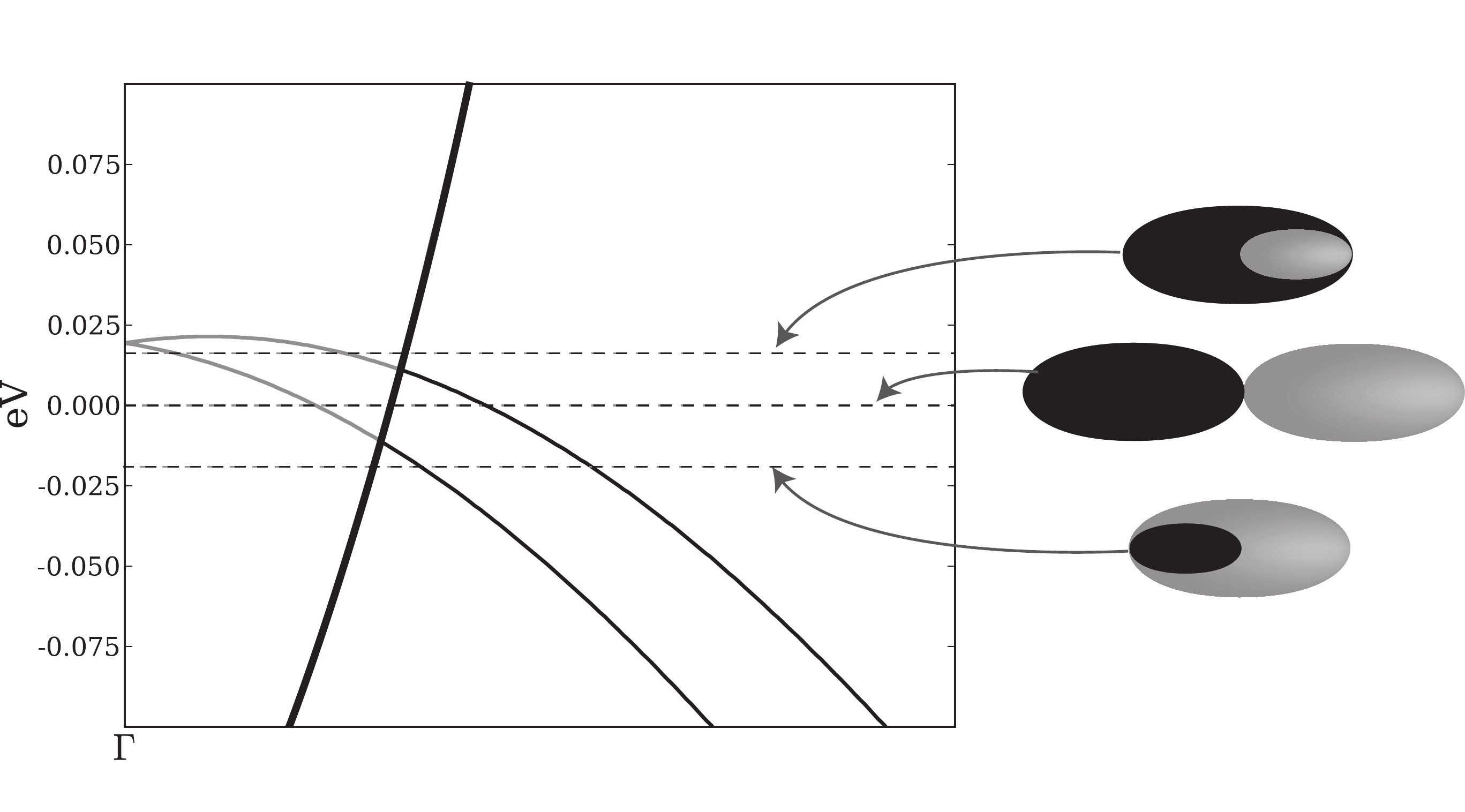}
}

\caption{{Band dispersion in tensile strained HgTe at the Fermi level as a function of momentum along the (111) axis. The twofold degenerate band (drawn as a thick line) is accidentally degenerate with both the nondegenerate bands (drawn as thin lines) within a very small energy range. Band dispersion around the accidental degeneracies is linear to leading order in the (111) direction. Right: Schematic illustration of the Fermi surface. The Fermi surface for various choices of Fermi energy (drawn in dashed lines) near the accidental degeneracies consists of two ellipsoids which contact at a point. This indicates that the bands disperse linearly in transverse directions as well around the accidental degeneracies.}}\label{fermisurf}
\end{figure}

\begin{figure}[t]
{
     \includegraphics [width=3in]{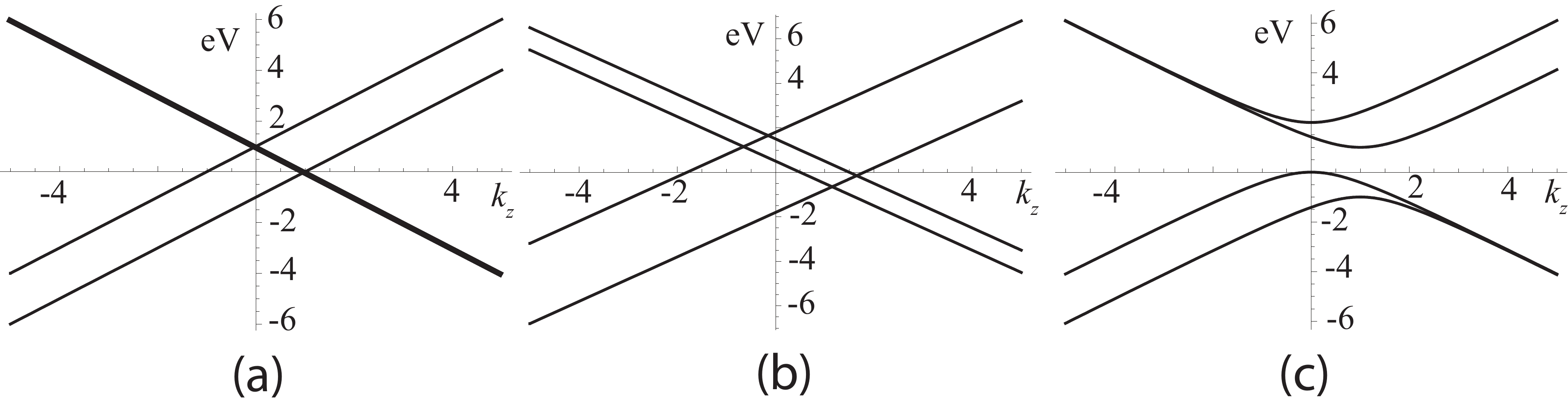}
}

\caption{{(a) Energy bands plotted as a function of momentum in the $k_z$ direction in the $\bf k\cdot p$ theory. There is a twofold degenerate band (thick line) intersecting two nondegenerate bands (thin lines) analogous to Fig.~\ref{fermisurf}. (b) Mirror and time reversal symmetry breaking perturbation proportional to $J_z$ splits the twofold degenerate band into two nondegenerate bands, giving rise to four Weyl points. (c) Rotation and mirror symmetry breaking perturbation proportional to $\tau_z\sigma_x$ gaps the system entirely.}}\label{kdotpbands}
\end{figure}

Figure~\ref{fermisurf} shows the band structure along the $(111)$ direction for $8\%$ tensile strain in HgTe. The set of momenta corresponding to a fixed energy form two ellipsoidal surfaces that contact at a point. This implies that the bands $\Lambda_4, \Lambda_5$ and $\Lambda_6$ disperse linearly in all directions around the accidental degeneracies. Therefore, under tensile strain the Fermi surface of HgTe grows from a point to a set of four ellipsoids, in two sets of Kramers paired momenta in the Brillouin zone. The two ellipsoids on one side of $\Gamma$ are connected to each other at the point ${\bf k}$ along the $(111)$ axis where the Fermi level crosses the band $\Lambda_6$. 

We note that $8\%$ tensile strain corresponds to $4\%$ strain in-plane, which is relatively large. This effect ensues as soon as a perturbation breaking the fourfold symmetry of zincblende turns on; there is no critical value of strain below which the unusual spin texture of the Fermi surface would disappear. While $8\%$ tensile strain, which corresponds to $4\%$ strain in-plane, is quite high, the ellipsoidal fermi surfaces ought to be observable at reduced strains; at $2\%$ tensile strain the energy difference between the threefold degeneracies is still 18meV, down from 25meV.  Even so, very high strains may be achievable through epitaxial growth of HgTe on an appropriate substrate~\cite{Faurie86p785}. 

\begin{figure*}
{
\subfigure[]{     \includegraphics [width=3in]{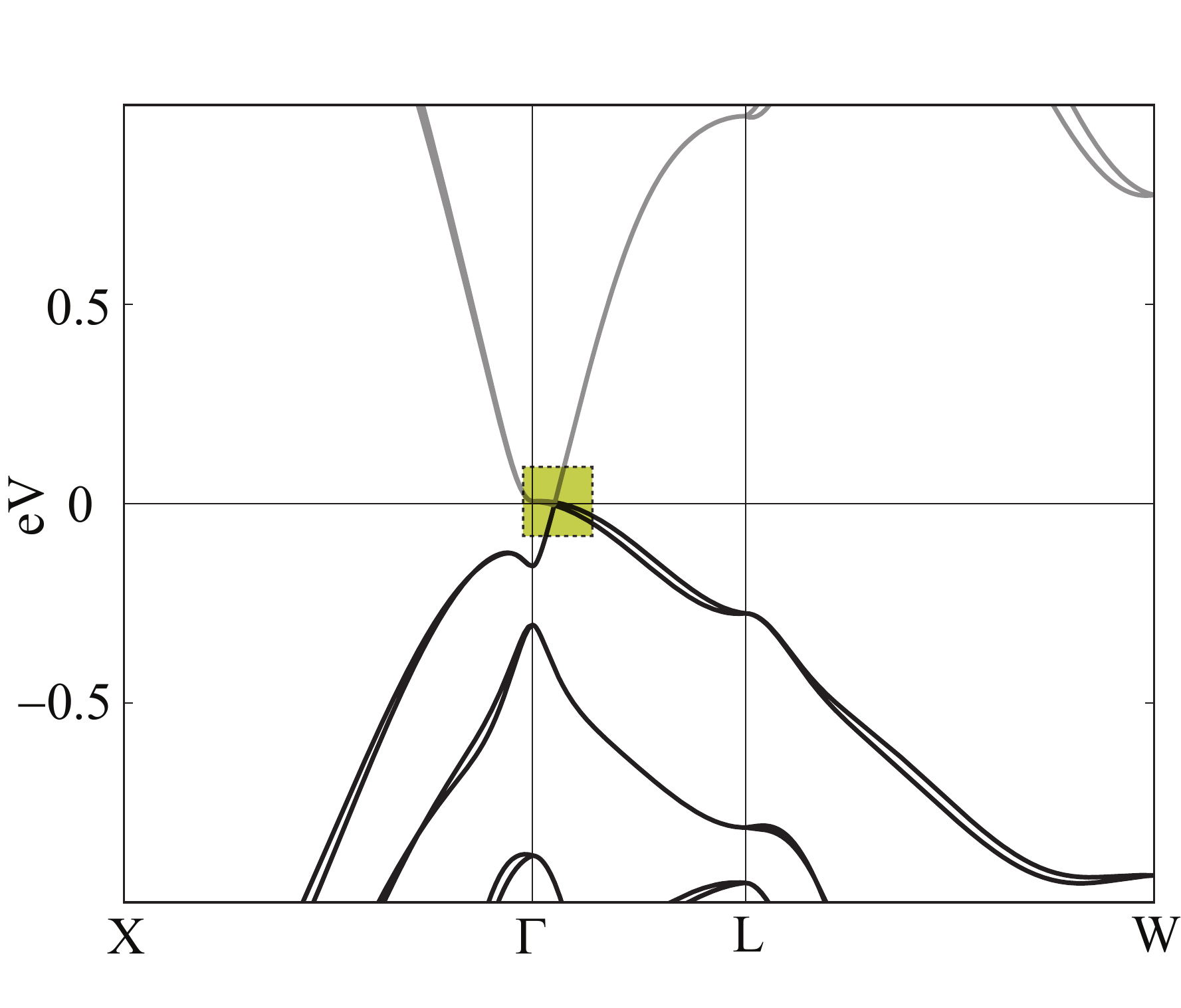}\label{fig:hgznte}}
\subfigure[]{     \includegraphics [width=3in]{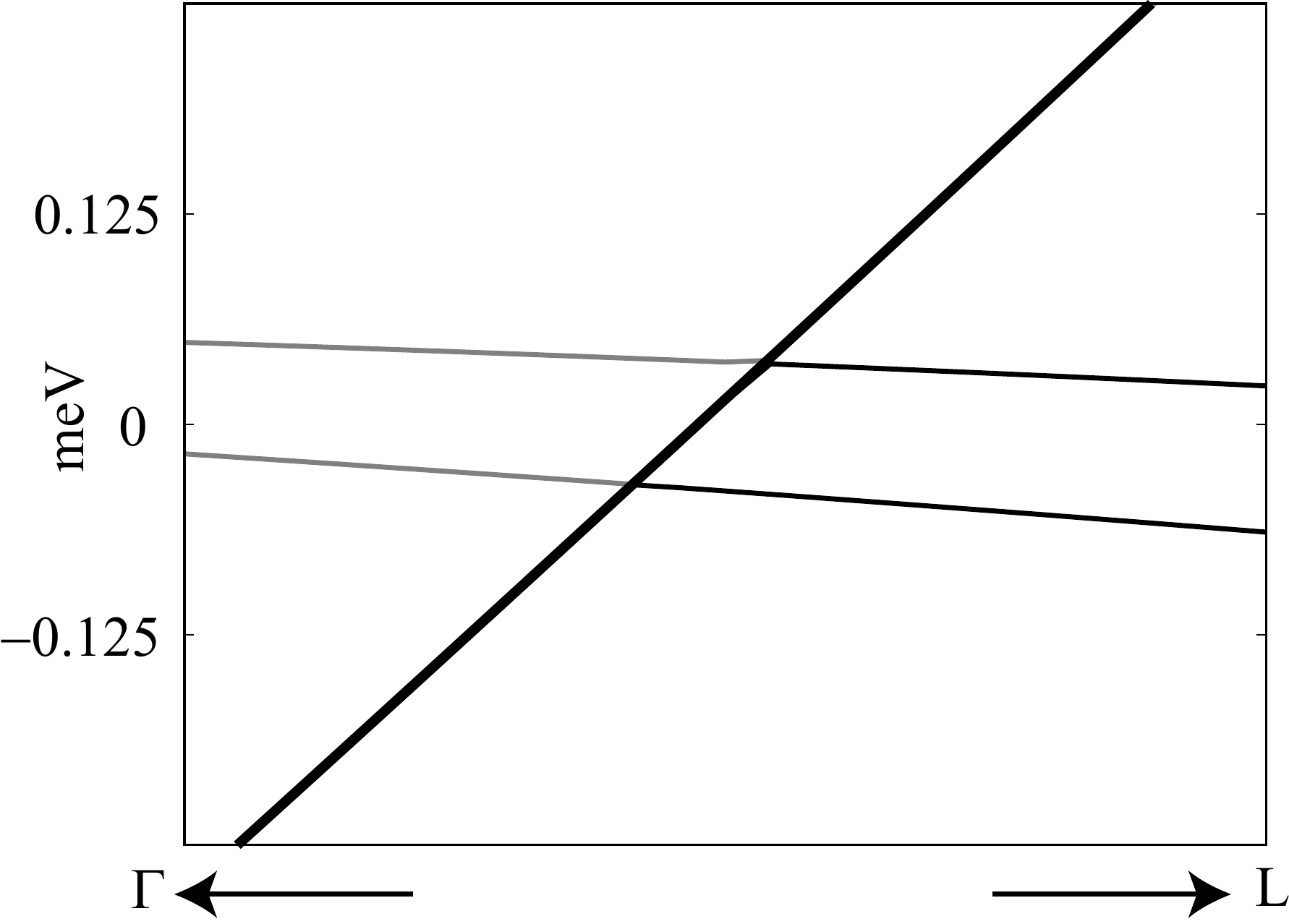}\label{fig:hgzntezoom}}
}
\caption{{~\subref{fig:hgznte} Band structure of $\rm Hg_{0.5}Zn_{0.5}Te$ is the same as tensile strained HgTe [Fig.~\ref{fig:pos}] except the shape and size of the Fermi surface and the Fermi velocities in the two materials. ~\subref{fig:hgzntezoom} Band dispersion at the Fermi level near the accidental degeneracies along the (111) axis. There is a twofold degenerate band (thick line) intersecting two nondegenerate bands (thin lines), essentially identical to tensile strained HgTe~[Fig.~\ref{fermisurf}].}}\label{hgzntebands}
\end{figure*}

Alternatively, strain can be introduced chemically.  Alloying HgTe with ZnTe introduces a size mismatch between Hg and Zn, causing both Zn and Hg atoms to move off center toward a Te atom; if all the Zn atoms move in the same direction, the symmetry breaking is the same as that created by tensile strain. A similar effect has also been reported in $\rm Cd_xZn_{1-x}Te$~\cite{Weil89p2744}, which shows rhombohedral distortions and Zn off-centering for a wide range of Zn concentration. Figure~\ref{hgzntebands} shows the band structure of ordered $\rm Hg_{0.5}Zn_{0.5}Te$, which is qualitatively the same as that of strained HgTe. The difference lies in the shape and size of the Fermi surface and the Fermi velocities in the two materials. As a caveat, it is important to note that DFT overestimates the magnitude of band inversion~\cite{Svane11p205205}; $\rm Hg_xCd_{1-x}Te$ for $x < 0.8$ is a band insulator~\cite{Scott69p4077}, and $\rm Hg_{0.5}Zn_{0.5}Te$ likely to be gapped in actuality. However, calculations for $\rm Hg_{0.75}Zn_{0.25}Te$ suggest that significant symmetry-breaking distortion persists for lower Zn concentration, consistent with the observations of similar behavior in $\rm Cd_{x}Zn_{1-x}Te$~\cite{Weil89p2744}. Therefore, if some ordered compound $\rm Hg_x Zn_{1-x} Te$ exists with $x$ such that distortion due to Zn substitution sufficiently splits the degeneracy at the Fermi level while still allowing band inversion, then it would possess an experimentally observable ellipsoidal Fermi surface. We assert that a time reversal and mirror symmetry breaking Zeeman field would split the degenerate band $\Lambda_6$ and turn $\rm Hg_x Zn_{1-x} Te$ into a Weyl semimetal. This is confirmed by our ${\bf k \cdot p}$ theory in Section~\ref{kdotptheory}. A Weyl semimetal obtained by a Zeeman perturbation to $\rm Hg_{x}Zn_{1-x}Te$ will have interesting phenomenological consequences such as Fermi-arc surface states~\cite{Vishwanath11p205101} and pressure induced anomalous Hall effect~\cite{Yang11p075129}.

While some experimental studies have been performed on Hg$_{1-x}$Zn$_{x}$Te, it is difficult to draw any direct conclusions without more detailed structural characterization of the samples~\cite{Sher86p2024,Berding87p3014}. We performed a phonon analysis of tensile strained HgTe and Hg$_{0.5}$Zn$_{0.5}$Te and found no structural instability. 


\section{${\bf K \cdot P}$ Theory}\label{kdotptheory}
We model the low energy physics of the states that span the representations $\Lambda_4, \Lambda_5$ and $\Lambda_6$ along the $(111)$ axis in terms of a four band ${\bf k \cdot p}$ theory around the accidental degeneracies. The states which belong to $\Lambda_6$ are $|p,\pm \frac{3}{2}\rangle$ whereas the ones belonging to $\Lambda_4$ and $\Lambda_5$ are $|p,\pm \frac{1}{2}\rangle$. The symmetry group is $C_{3v}=\{1, x,x^2,y, xy, x^2y\}$ with the group multiplication rules $x^3= y^2=1, yx=x^2y$ in addition to time reversal symmetry. Here $x$ represents a threefold rotation symmetry while $y$ represents a mirror symmetry. The accidental degeneracies occur at points ${\bf k}_0 \not = 0$ along the line from $\Gamma$ to $\rm L$, which are not time reversal symmetric. So the time reversal operator $\Theta$ maps the $\bf k\cdot p$ Hamiltonian around ${\bf k}_0$ to its time reversed partner around $-{\bf k}_0$. 

We choose an angular momentum basis for the $\bf k \cdot p$ theory. Since we are dealing with four half-integer spin states, we require a double valued representation of $C_{3v}$, such that $x$ and $y$ take the form of matrices $\hat{R}$ and $\hat{M}$ belonging to SU(4) that satisfy $\hat{R}^3= \hat{M}^2=-1$. These operators satisfy the group multiplication rules and are consistent with the interpretation of the operators $\hat{M}, \hat{R}\hat{M}, \hat{R}^2\hat{M}$ as mirror symmetries (all of which square to −1) for particles with half integer spin.

Orienting the $k_z$ direction along the (111) axis, the threefold rotation operator is diagonal and can be written as $\hat{R} = \exp(i\hat{J}_z 2\pi/3 )$ where $\hat{J}_z$ is the angular momentum operator. The mirror operator in this basis takes the form $\hat{M} = i\tau_y\otimes \sigma_x$. $\vec{\tau}$ and $\vec{\sigma}$ denote Pauli matrices. Tensor products of Pauli matrices can be used as basis vectors for $4\times 4$ hermitian matrices. Neither $\vec{\tau}$ nor $\vec{\sigma}$ \emph{individually} corresponds to a real or pseudo spin degree of freedom. In what follows, we will write $\tau_i\otimes \sigma_j$ as $\tau_i\sigma_j$ where the tensor product is implied. In the absence of inversion symmetry $\mathcal{P}$, the local antiunitary operator $\mathcal{P}\Theta$ is not a symmetry. $\mathcal{P}\Theta $ is a local operator because both $\mathcal{P}$ and $\Theta$ send $\bf k$ to $\bf -k$. In the angular momentum basis, it can be written as $\mathcal{P}\Theta = i\tau_x\sigma_y K$, where $K$ is the complex conjugation operator.

A Hamiltonian linear in $\bf k$ which respects these symmetries is,
\begin{flalign}
\hat{H}({\bf k})= k_x\tau_x+ k_y\tau_y +k_z\tau_z\sigma_z + \kappa \label{kdotp}
\end{flalign}
where
\begin{flalign*}
	\kappa=\left(\begin{array}{cccc}
      0  &  0  &  0  & -i   \\ 
      0  &  1  &  0  &  0   \\
      0  &  0  &  1  &  0   \\
      i  &  0  &  0  &  0   
	\end{array}\right)
\end{flalign*}
is chosen to break the local antiunitary symmetry $\mathcal{P}\Theta$. With this choice of $\kappa$, time reversal symmetry is preserved while inversion is broken if we require that $\Theta \hat{H}({\bf k}) \Theta^{-1} = \tau_x k_x + \tau_y k_y +\tau_z \sigma_z k_z + \kappa^T $ is the effective ${\bf k\cdot p}$ Hamiltonian around the time reversed location of the accidental degeneracies. Such a constraint is allowed because the ${\bf k \cdot p}$ theory is localized around a Brillouin zone momentum ${\bf k}_0\not = 0$ along the $(111)$ axis.

Figure~\ref{kdotpbands} shows the energy spectrum of the Hamiltonian in Eq.~(\ref{kdotp}) along the $k_z$ axis. The degenerate band $\Lambda_6$ splits to give four Weyl points under a mirror and time reversal symmetry breaking Zeeman term proportional to $\sigma_z$ or $J_z$ where $J_z$ is the angular momentum operator of $p_{3/2}$ states. Fig.~\ref{kdotpbands}(b) shows the splitting of $\Lambda_6$ to give four Weyl points due to a perturbation proportional to $J_z$. A rotation and mirror symmetry breaking perturbation such as $\tau_z \sigma_x$ gaps the system entirely~[Fig.~\ref{kdotpbands}(c)].

It is possible to split $\Lambda_6$ into two parallel bands as in Fig.~\ref{kdotpbands}(b) by a mirror preserving term $\tau_y \sigma_x$ which breaks time reversal and rotation symmetry. This however does not lead to Weyl points; even though the degeneracies involve two states, the bands develop a quadratic dispersion in one direction. Mirror symmetry about a plane constrains all points $\bf k$ on the plane to have zero Chern number. This is explained as follows: The Chern number of a certain point $\bf k$ in the Brillouin zone is given by an integer $n = 1/(2\pi i) \oint_{S} \tr \mathcal{\vec{F}}\cdot\hat{n}$ where $\mathcal{\vec{F}}_{ij} = \nabla_{\bf k} \times \langle \psi_{i}| \nabla_{\bf k} |\psi_j\rangle$ is the Berry curvature, $\hat{n}$ is a unit vector normal to the surface $S$ that contains $\bf k$, and the trace includes only valence states $|\psi_{i}\rangle, i=1,..N$. A mirror symmetry about a plane that contains $\bf k$ reverses the orientation of the surface $S$ while the integrand is the same for opposing points on $S$. Hence the Chern number of all points on a mirror plane must be zero.  Therefore, as long as mirror symmetry is present, tensile strained HgTe cannot host Weyl points along the $(111)$ axis.

\section{Spin Texture}\label{abinitio}
Since the dispersion of bands around the accidental degeneracies in Figs.~\ref{fermisurf} and~\ref{kdotpbands}(a) is linear in ${\bf k}$, it is natural to ask if the accidental degeneracies have an associated Chern number. However since the (111) axis belongs to a mirror plane, the Chern number of all points on that axis should be zero. In other words, the surface integral of the normal component of the spin texture over a closed surface enclosing the accidental degeneracies will be zero. The ellipsoidal Fermi surface in $\rm Hg_{0.5}Zn_{0.5}Te$ also encloses the accidental degeneracies, as shown in Fig.~\ref{fermisurf}. At the ${\bf k \cdot p}$ level, this corresponds to two spheres in $\bf k$ space parametrized as, $S_1:~\mathcal{E}^2= k_x^2+k_y^2+(1-k_z)^2$ and $S_2:~(\mathcal{E}-1)^2= k_x^2+k_y^2+k_z^2$ where $\mathcal{E}$ is the Fermi energy. The average spin field is given by $\langle \psi_i| \hat{{\bf S}}|\psi_i\rangle$ where $|\psi_i\rangle$ is a valence state on one of the Fermi ellipsoids parametrized by $\mathcal{E}$ and $\hat{{\bf S}}$ is the spin operator for the $ p_{3/2}$ states. In the $\bf k \cdot p$ theory, this evaluates to,
 
\begin{flalign}
\langle\hat{\mathbf{S}}\rangle_i&=\alpha_i(\cE, k_z) \left[\sin\left(\theta\right)\hat{\mathbf{i}}+\cos\left(\theta\right)\hat{\mathbf{j}}\right]\nonumber \\
&+\beta_i(\cE, k_z)\left[\sin\left(2\theta\right)\hat{\mathbf{i}}-\cos\left(2\theta\right)\hat{\mathbf{j}}\right]\label{spin}
\end{flalign}
where the subscript $i=1,2$ identifies one of the Fermi ellipsoids $S_i$ and $\theta$ is the azimuthal angle in the $k_xk_y$ plane that parametrizes the ellipsoids. The coefficients $\alpha_i, \beta_i$ are,
\begin{flalign*}
\alpha_1(\cE, k_z)&=-\sqrt{\frac{4(\mathcal{E}+1-k_z)}{3(\mathcal{E}-1+k_z)}}~, \beta_1(\cE, k_z) = -\frac{\alpha_1(\cE,k_z)^2}{2}\\
\alpha_2(\cE, k_z)&=\sqrt{\frac{4(\mathcal{E}-1-k_z)}{3(\mathcal{E}-1+k_z)}}~, \beta_2(\cE, k_z) = \frac{\alpha_2(\cE,k_z)^2}{2}\\
\end{flalign*}
\begin{figure*}
{
     \includegraphics[width=6in]{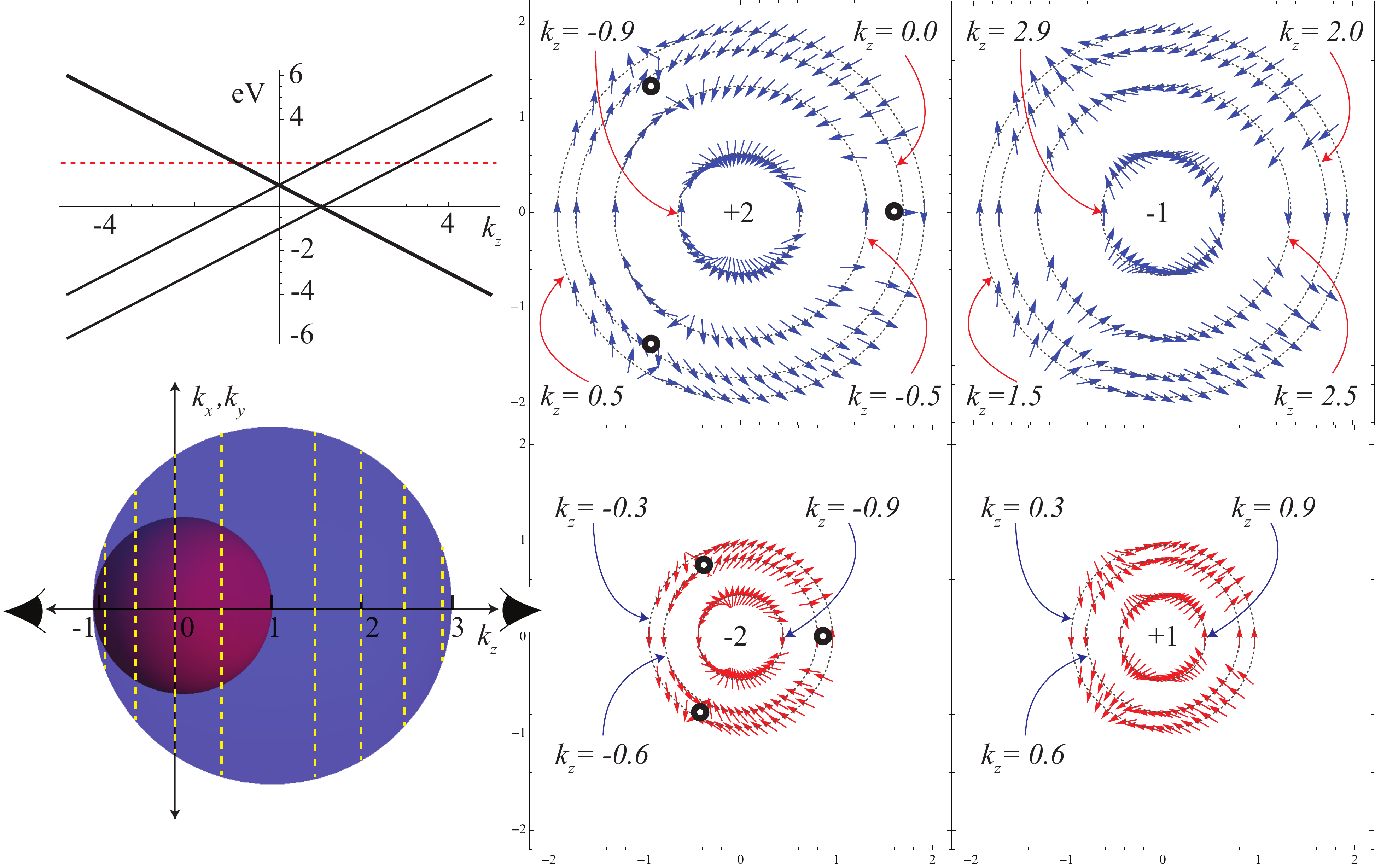}
}
\caption{{Spin texture on the Fermi surface of tensile strained HgTe as predicted by the $\bf k\cdot p$ theory of Eq.~\ref{kdotp}. Top Left: Band dispersion in the $\bf k\cdot p$ theory as a function of momentum along the (111) axis. The Fermi level $\mathcal{E}_F$(dashed line) is set to be 2 eV in the units of Eq.~\ref{kdotp}. Bottom Left: The Fermi surface for $\mathcal{E}_F = 2 \rm~eV$ at the $\bf k\cdot p$ level consists of two spherical shells which touch at the point $k_z = -1$. Top Right: The spin texture of valence states on the blue Fermi spherical shell as a function of momentum in the $k_x, k_y$ plane. Each circle corresponds to a constant $k_z$ (indicated by dashed yellow lines cutting through the Fermi surface on the left). Bottom Right: The spin texture of valence states on the red Fermi spherical shell as a function of momentum in the $k_x, k_y$ plane. The spins are locked in the $k_x,k_y$ plane for both Fermi shells. The winding number of spins on each Fermi shell changes as one moves from one end along $k_z$ to the other. There are three points indicated in yellow where the spin texture vanishes. The spin winding number flips from $\pm 2$ to $\mp 1$ at these spin singular points.}}\label{spintexture}
\end{figure*}
The three mirror planes of the group $C_{3v}$, all of which contain the $k_z$ axis, lock the spin expectation values in the $k_xk_y$ plane. Figure~\ref{spintexture} shows how the spin texture evolves as a function of $k_z$ around the Fermi ellipsoids. For fixed $\cE$, $k_z$ ranges between $1-|\cE|$ and $1+|\cE|$ on $S_1$ and between $-|1-\cE|$ and $|1-\cE|$ on $S_2$. When $k_z$ is close to $1-\cE$ on either ellipsoid, the term with the coefficient $\beta_i$ dominates and the spin winding number is $-2$, whereas when $k_z$ is close to $\cE+1$ on $S_1$ or $\cE-1$ on $S_2$ the term with the coefficent $\alpha_i$ dominates, and the spin winding number is $+1$. Therefore, the spin winding number must change abruptly from one end of the Fermi ellipsoids to the other, since the spins cannot develop a component in the $k_z$ direction due to mirror symmetry. Hence, there must exist singular points on the Fermi ellipsoids where the spin expectation values vanish. Figure~\ref{spintexture} shows the points on the Fermi ellipsoids where the spin expectation value goes to zero. There are three such singular points which are related to each other by threefold symmetry, and their exact location can be calculated in the ${\bf k \cdot p}$ theory. Note that although the $\alpha_i$ diverge in the limit $k_z \to 1-\cE$, the points where $k_z=1-\cE$ correspond to the degenerate band $\Lambda_6$, and since $\alpha_1, \beta_1$ and $\alpha_2, \beta_2$ have opposite signs, the total spin $\langle\hat{\mathbf{S}}\rangle_1 +\langle\hat{\mathbf{S}}\rangle_2=0$ for states in $\Lambda_6$. 
\begin{figure*}[t]
{
\subfigure[]{     \includegraphics [width=3in]{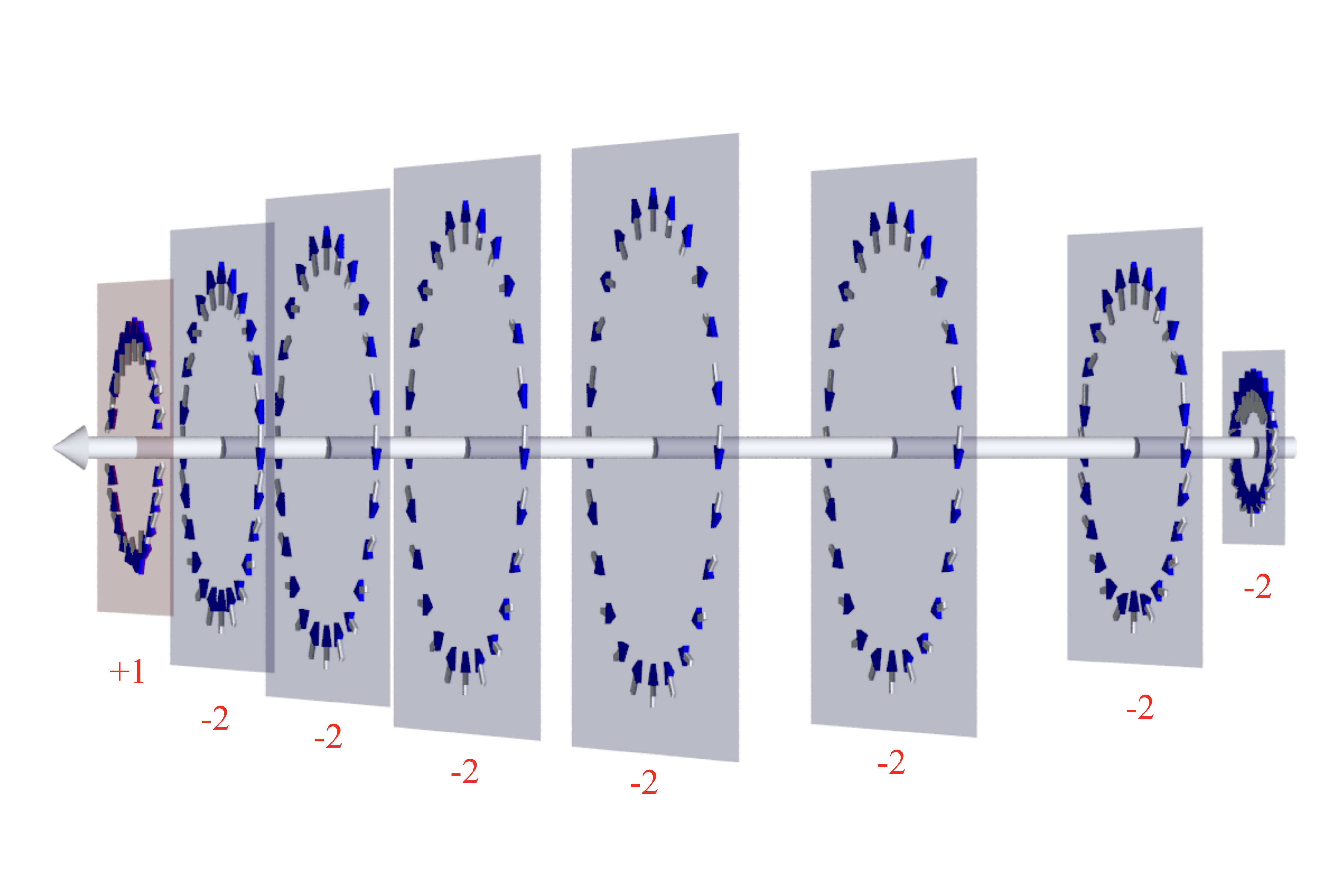}\label{fig:spinall}}
\subfigure[]{     \includegraphics [width=3in]{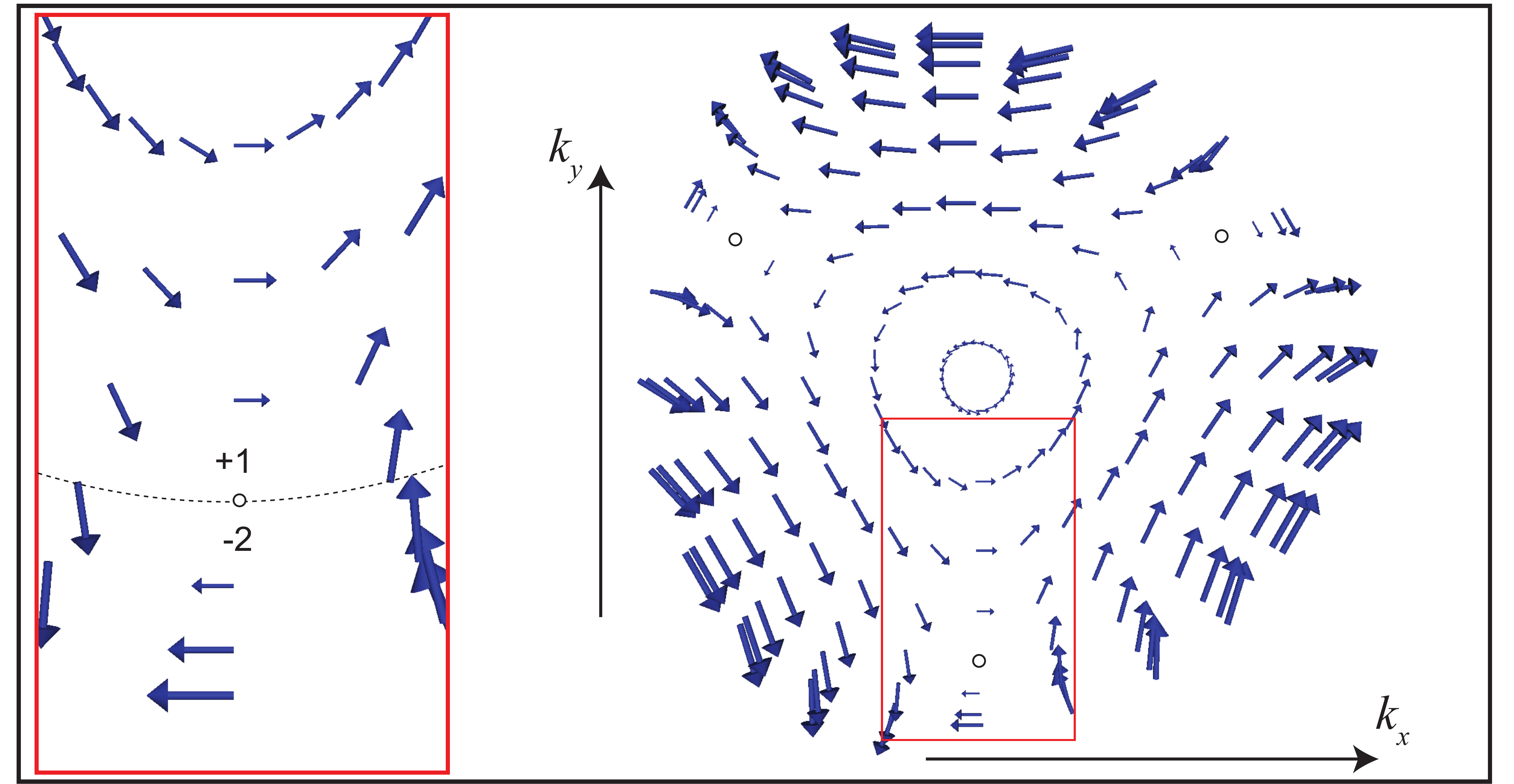}\label{fig:cap}}
}
\caption{{Spin texture on the Fermi surface of $\rm Hg_{0.5}Zn_{0.5}Te$ at the \emph{ab initio} DFT level. ~\subref{fig:spinall} Spins on a Fermi surface ellipsoid of Hg$_{0.5}$Zn$_{0.5}$Te are locked in the plane perpendicular to the (111) axis. The spin winding number around the (111) axis changes from +1 to -2 from the left end of the ellipsoid to the right end. ~\subref{fig:cap} Spin texture in the two dimensional plane transverse to the (111) direction. Each concentric loop of spins corresponds to a fixed value of momentum along the (111) axis. The spin winding number around the (111) axis changes from the inner circles to the outer circles on the Fermi ellipsoid. Since spins are locked on the transverse plane, their winding number changes as they vanish at three rotation symmetry related singular points (drawn as white circles outlined in black). Left: A higher resolution image showing the change in the spin winding number from +1 to -2 as we move from the inner radii to the outer radii on the right. For visual clarity, the magnitude of all vectors is normalized in~\subref{fig:spinall}, while in ~\subref{fig:cap} the size of each vector equals the square root of its magnitude.}}\label{spinsurf}
\end{figure*}

We have carried out \emph{ab initio} calculations to confirm that tensile strained HgTe and $\rm Hg_{0.5}Zn_{0.5} Te$ exhibit identical spin texture at the Fermi surface. Figure~\ref{spinsurf} illustrates the spin texture on the Fermi surface of $\rm Hg_{0.5}Zn_{0.5} Te$. The spins are locked in a two dimensional plane perpendicular to the (111) axis, and their winding number changes from one end of the Fermi ellipsoid to the other. Furthermore, the abrupt change in the winding number is accomodated by the vanishing of the spin expectation value at three rotation symmetry related points on the Fermi ellipsoid. Density functional theory calculations were performed in the {\footnotesize QUANTUM ESPRESSO} suite using the Perdew-Burke-Ernzerhof-type generalized gradient approximation (GGA)~\cite{Giannozzi09p395502}. All calculations used an $8\times8\times8$ Monkhorst-Pack $\rm \bf k$-point mesh with a plane-wave cutoff of 50 Ry. The pseudopotentials representing the atoms in these simulations were generated by the {\footnotesize OPIUM} package; the pseudopotentials were norm-conserving and optimized, and included the full relativistic correction~\cite{Rappe90p1227, Ramer99p12471}.  

We have shown that HgTe tensile strained in the (111) axis is semimetallic and exhibits a Fermi surface consisting of two ellipsoids that contact at a point. Due to mirror and threefold symmetry of the zincblende lattice, the spin texture on the Fermi ellipsoids is locked in a two dimensional plane and vanishes at special singular points. We have derived a $\bf k\cdot p$ theory to understand this phenomenon at a qualitative level. We propose that this symmetry protected spin texture can be observed in $\rm Hg_{1-x}Zn_xTe$ for values of $x$ small enough to allow band inversion while sufficiently breaking symmetry.

Note: We recently became aware that the authors of Reference~\onlinecite{Soluyanov11p235401} studied the effects of biaxial strain (referred to as `strain' in that paper) along the (111) axis in HgTe. They found that the fourfold degeneracy at $\Gamma$ was lifted and no accidental degeneracies resulted for either positive and negative strain. Therefore, biaxial strain (compressive or tensile) is similar to compressive uniaxial strain in terms of its effect on band structure. In contrast, we find that uniaxial tensile strain not only maintains the accidental degeneracies already present in HgTe but shifts them further away from $\Gamma$.

\emph{Acknowledgments}. We thank David Vanderbilt and Alexey A. Soluyanov for helpful discussions. This work was supported in part by the MRSEC program of the National Science Foundation under Grant No. DMR11-20901 (S. M. Y.),  by the Department of Energy under Grant No. FG02-ER45118 (E. J. M. and S. Z.), and by the National Science Foundation under Grant No. DMR11-24696 (A. M. R.) and Grant No. DMR09-06175 (C. L. K. and J. C. Y. T.). D. C. was supported by the REU Program at LRSM, University of Pennsylvania. S. M. Y. acknowledges computational support from the High Performance Computing Modernization Office.

\end{document}